\documentclass[12pt,epsf]{article}
\usepackage{epsf,graphicx} 
\newcommand{\bb}{\begin{eqnarray}}
\newcommand{\ee}{\end{eqnarray}}
\begin{document}
\begin{center}
{{\Large {\bf
Thermodynamics of a class of Kerr-Bolt-Ads spacetime.}}}
\vskip 2cm
Tanwi Ghosh$^a$\footnote{e-mail: tanwi@theory.saha.ernet.in},
 
\vskip 1cm
$^a$Saha Institute of Nuclear Physics, Bidhannagar, Kolkata
700 064, INDIA\\

\end{center}
 
\vskip 1cm
\
\begin{abstract}
Using a boundary prescription motivated by the Ads-Cft conjecture, I study the thermodynamical properties of
the class of Kerr-Bolt-Ads spacetime.The stability conditions and the complete phase diagrams are investigated.
\end{abstract}
\newpage

{\bf{\Large{Introduction}}}\\
 
 The thermodynamical properties of gravity have been  connected with the presence of blackholes [1,2].A physical entropy S and temperature $1/\beta$ can be ascribed to a given blackhole configuration,where these quanties are respectively proportional to the area and surface gravity of the event horizon.The conjecture equivalence of string theory on anti-de-Sitter (Ads) spaces (times some compact manifold ) and certain superconformal gauge theories living on the boundary of Ads[3-6 ] has lead to an increasing interest in asymptotically Anti-de Sitter black-holes.There are two known bulk solutions with boundary $ {S^1}\times{S^P} $ (where ${ S^1 }$ is the Euclidean time ).The more obvious one is Ads itself while the other is Euclidean Ads solutions. The former one governs low temperature phase while the latter controls the high
temperature phase.
Thermodynamics of Ads black holes [7-9] and also of Schwarschild-Ads black-holes were discovered by Hawking and Page [10 ],where a phase transition from thermal Ads space to
 a black-hole phase was occured as temperature increases.At a certain temperature thermal radiation in Ads space becomes unstable, and 
eventually collapses to a black-hole.The thermodynamics of R-charged black-holes in four,five and seven dimensions has been studied in
 detail in [11 ].Recently thermodynamics properties of rotating black-holes in Ads spaces has been studied [12-16].It has been 
demonstrated recently that the entropy can be associated with a gravitational system, one containing Misner string[17-19 ].These objects are the
 gravitational analogue of Dirac strings, and arise whenever the gravitational field in the Euclidian regime has a U(1) isometry group
 ( generated by a Killing vector $\zeta$ which is timelike in the Lorengian regime) with a fixed point set of co-dimension $d_f<d-2$
 ( called a Nut [ 20] ). A space-time can also contain both black-holes ( for which $d_f>d-2$-called a Bolt ) and Misner string.In this
 space-time the expressions for entropy is given by,
$S=\beta H_\infty-(I_v+I_b+I_{ct})$ where $H_\infty=M+\Omega J$ with $M=Q[\partial/\partial \tau]$ and $J=Q[\partial/\partial \phi]$ being
 the conserved charges associated with the Killing vectors $\partial/\partial \tau$ and $\partial/\partial\phi$ where $\Omega=\frac{a}{r_+^2-a^2-N^2}$.$I_v$,$I_b$ and $I_{ct}$ are volume term,boundary term and counter term of action [19].
The class of Kerr-NUT-Ads spacetime has the metric form
\begin{eqnarray}
\nonumber ds^2 &=& \frac{v(r)\left(d\tau - \left[2N cos(\theta)- a sin^2(\theta) \right]d\phi \right)^2 +\mathcal{H}(\theta)sin^2\theta 
       \left[ad\tau -\left(r^2-N^2-a^2\right)d\phi\right]^2}{\chi^4\left(r^2-\left[N+acos(\theta)\right]^2\right)}\\
     &&  +\left(r^2-\left[N+acos(\theta)\right]^2\right)\left(\frac{dr^2}{v(r)}+\frac{d(\theta)^2}{\mathcal{H}(\theta)}\right)
\end{eqnarray}
where the Einstein field equations imply
\begin{eqnarray}
\nonumber \mathcal{H}=1+\frac{qN^2}{l^2}+\frac{\left[2N+acos(\theta)\right]^2}{l^2}
\end {eqnarray}
\begin{eqnarray}
\nonumber v(r)=\frac{r^4}{l^2}+\frac{\left[(q-2)N^2-a^2+l^2\right]r^2}{l^2}-2mr-\frac{(a+N)(a-N)(qN^2+l^2+N^2)}{l^2}
\end{eqnarray}
and where the periodicity in $\tau$ and the parameters q and $\chi$ are chosen so that conical singularities are avoided.
There are string singularities at both the north and the south poles in this metric when$ N\neq 0$ .
 In the $(\theta ,\phi )$ section these considerations may be implemented as follows. Writing the metric as
\begin{eqnarray}
\nonumber ds^2 &=& g_{\tau\tau}\left(d\tau+\frac{g_{\tau\phi}d\phi}{g_{\tau\tau}}\right)^2+g_{rr}dr^2+g_{\theta\theta}d\theta^2\\
            &&   +\left(g_{\phi\phi}-\frac{g_{\tau\phi}^2}{g_{\tau\tau}}\right)d\phi^2
\end{eqnarray}

conical singularities in the $(\theta ,\phi)$ section will be absent provided the metric in this section is conformal to
$ d\theta^2+\theta^2d\phi^2 $ near $\theta=0$,and to$d\theta^2+(\theta-\pi)^2d\phi^2$near $\theta=\pi$.Expanding the $(\theta,\phi)$part of the metric about these respective points yields$\chi^2=\mathcal{H}(0)$and $\chi^2=\mathcal{H}(\pi)$.
   
For $N\neq 0$   these relations can not be simultaneously satisfied. However for the form of the metric given above 
there are string singularities at each of these values of $\theta$ when $N\neq 0$ , so this is a moot point.Transforming $\tau\rightarrow\tau\pm2N\phi$  respectively removes
the string singularities at $\theta=0,\pi$; one can then use two nonsingular coordinate patches at each of the poles and then match them elsewhere via a simple coordinate transformations.
$\chi=\sqrt{1+a^2/l^2}$     can be taken so that conical singularities are manifestly removed when N = 0 . Further requiring that the conformal
 factor be unity when $a=0$ yields $q=-4$.
The periodicity in $\tau$ is more subtle. The locations of the nut is at$ r=\sqrt{a^2+N^2}\equiv r_N $, where the area of surfaces orthogonal to the (r,$\tau$ )
section vanishes. The Misner string singularity runs along the positive and negative z-axis then implies that$\tau$  also has 
period  $\frac{2\pi}{\kappa}$ where $ \kappa=\sqrt{-\nabla_{\mu}\zeta_{\nu}\nabla^\mu\zeta^\nu/2}$  , with $ \zeta =\frac{\partial}{\partial\tau}+\Omega(\partial/\partial\phi)$ ,being the killing field normal to the horizon, and where$\Omega=\frac{a}{r_+^2-(a+N)^2}$  is the angular velocity of the horizon.
Explicitly$\kappa=\frac{\acute{v}(r_+)}{2\chi^2(r_+^2-r_N^2)}$ where $ v(r_+ )=0 $, being the locations of the foliation breakdown.If $r_+=r_N $  ,in this case regularity of the solutions demands
 that v(r) have a double root there. However this requirement turns out to be incompatible with the $ \tau $ periodicity constraint unless a=0.
Hence $ r_+> r_N $  , and there are no regular Kerr-Nut or Kerr-Nut-Ads solutions [20].
Denoting the unit radial normal to $\partial M $ (with induced metric$\gamma_{\mu\nu}$ )by$n^{\nu}$  , the quasilocal mass and angular-momenta are given by the expressions
\begin{eqnarray}
\nonumber M=\int d^2x\sqrt{\sigma}\zeta^\nu(ku_\nu+j_\nu)
\end{eqnarray}
\begin{eqnarray}
\nonumber J=\int d^2x\sqrt{\sigma}\psi^\nu(ku_\nu+j_\nu)
\end{eqnarray}
where $\zeta=\frac{\partial}{\partial\tau}$,$\psi=\frac{\partial}{\partial\phi}$ and k is the trace of extensive curvature.$k_{\mu\nu}=\sigma_\mu^\alpha\sigma_\nu^\beta\nabla_\alpha n_\beta$ is the extrinsic curvature of the 2-boundary which is the intersection of $\partial M$and 
$\Sigma_\tau$,with metric $\sigma_{\mu\nu}=\gamma_{\mu\nu}-n_{\mu}n_{\nu} $.The vector $j_\nu$ is $ j_\nu=\sigma_\nu^\beta n^\alpha \nabla_\beta u_\alpha$ is the angular momentum vector of the 2-boundary.From [19] $M=\frac{m}{\chi^4}$, $J=\frac{ma}{\chi^4}$ for each of the Kerr,Kerr-Ads,Kerr-Bolt and Kerr-Bolt-Ads solutions. The parameter m obeys the constraints v($r_+$)=0
.
In this Kerr-Bolt-Ads space we would like to study the thermodynamical properties of black-holes.\\

{\bf\large { Thermodynamics Of Kerr-Bolt-Ads Spacetime}}:\\

In Kerr-Bolt-Ads spacetime, the action and entropies are [19] given by
\begin{eqnarray}
\nonumber I=-\pi\frac{(r_+^2-a^2-N^2)\left[r_+^4-(a^2+l^2)r_+^2+(N^2-a^2)(3N^2-l^2)\right]}{\left[3r_+^4+(l^2-a^2-6N^2)r_+^2+(N^2-a^2)(3N^2-l^2)\right]\chi^2}
\end{eqnarray}
\begin{eqnarray}
\nonumber S=\pi\frac{3r_+^6+(l^2-4a^2-15N^2)r_+^4+(a^2+3N^2)^2r_+^2+(N^2-a^2)^2(3N^2-l^2)}{\left[3r_+^4+(l^2-a^2-6N^2)r_+2+(N^2-a^2)(3N^2-l^2)\right]\chi^2}
\end{eqnarray}
Entropy is not one-quarter of the area,due to the presence of the Misner string.Entropy is not positive definite at a certain range of 
parameters N.A negative entropy certainly would appear to be a sign of pathological bahaviour. It may be a consequence of the particular
choice of time slicing.S is plotted as a function of N in Fig1.

Using the expressions for S,the angular velocity of this spacetime is given by
\begin{eqnarray}
\nonumber \Omega=J\frac{\left[3r_+^4M^2+(l^2M^2-J^2-6M^2N^2)r_+^2+(N^2M^2-J^2)(3N^2-l^2)\right](M^2+J^2/l^2)}{M\left[3r_+^6M^4+(l^2M^4-4J^2M^2-15N^2M^4)r_+^4+(J^2+3N^2M^2)^2r_+^2+(N^2M^2-J^2)^2(3N^2-l^2)\right]}+\frac{J}{Ml^2}
\end{eqnarray}
Where
\begin{eqnarray}
\nonumber \Omega_H=J\frac{\left[3r_+^4M^2+(l^2M^2-J^2-6M^2N^2)r_+^2+(N^2M^2-J^2)(3N^2-l^2)\right](M^2+J^2/l^2)}{M\left[3r_+^6M^4+(l^2M^4
-4J^2M^2-15N^2M^4)r_+^4+(J^2+3N^2M^2)^2r_+^2+(N^2M^2-J^2)^2(3N^2-l^2)\right]}
\end{eqnarray}
the angular velocity of the black-hole at the event horizon,and $\Omega_\infty=\frac{J}{Ml^2}$ is the angular velocity at infinity.

  We have plotted angular velocity as a function of angular momentum J in Fig2 and Fig3 .

From these curves we noticed that $\Omega $ is not positive 
for large M and it almost vanishes for small M. This behaviour 
is very similar to Kerr-Ads case [21].\\

Using the expressions for $\kappa$,the temperature of this spacetime is 
\begin{eqnarray}
\nonumber \frac{1}{T}=\beta=\frac{2\pi}{\kappa}=2\pi\frac{2\chi^2(r_+^2-a^2-N^2)}{\acute{v}(r_+)}=\frac{4\pi\chi^2(r_+^2-a^2-N^2)}{\left[3r_+^3/l^2+r_+/l^2(l^2-a^2-6N^2)+\frac{(a-N)(a+N)(l^2-3N^2)}{r_+l^2}\right]}
\end {eqnarray}
The specific heat $C_{J,N}$of this spacetime can be evaluated by

\begin{eqnarray}
C_{J,N}=\left.\frac{\partial M}{\partial T}\right|_{J,N}=A/B
\end{eqnarray} 
where

\begin{eqnarray}
A=8\pi l^2\left[M^2/l^2(3r_+^2/2+(l^2-6N^2)/2-\frac{N^2(l^2-3N^2)}{2r_+^2})-J^2/(2l^2)+\frac{J^2}{2r_+^2}-\frac{3J^2N^2}{2r_+^2l^2}\right]
\end{eqnarray}
\begin{eqnarray}
\nonumber B&=&\left[12r_+^3M^4+2r_+(l^2M^4-6N^2M^4-J^2M^2)-8\pi T l^2\left(\frac{J^2}{l^2}(3r_+^2M^2-M^2N^2-J^2) \right.\right. \\
            && \left.\left.  + (3r_+^2M^4-N^2M^4-M^2J^2)\right)\right] \\
            &&\times \left[3M^2+2J^2/l^2-\frac{J^4}{(M^2) (l^4)}-M/l^2\left(r_+^3+r_+(l^2-3N^2)+\frac{N^2(l^2-3N^2)}{r_+}\right)\right] \\
            &&+\left[12M^3r_+^4+r_+^2(4l^2M^3-2MJ^2-24N^2M^3)+ \right. \\
            && \;\; 2MJ^2(l^2-3N^2)-4M^3N^2(l^2-3N^2) \\
            &&-\left. 8\pi T l^2(J^2/l^2(2Mr_+^3-2Mr_+N^2)+(4r_+^3M^3-4r_+N^2M^3-2r_+MJ^2))\right] \\
            &&\times \left[M^2/l^2\left(3r_+^2/2+(l^2-6N^2)/2-N^2\frac{(l^2-3N^2)}{2r_+^2})\right)
             -\frac{J^2}{2l^2}+\frac{J^2}{2r_+^2}-\frac{3J^2N^2}{2l^2r_+^2}\right] \\
\end{eqnarray}

In order to study the thermal stability of this spacetime,it is convenient to examine specific heat.In the following curves $ C_{J,N}$
is plotted as a function of N.\\

From Fig4 it is found that for small M and low temperature $ C_{J,N}$ is not positive in this range of N. Fig5 shows that for large M and
 high temperature $ C_{J,N}$ approaches its  positive value after a certain range..As from Fig3 and Fig4 , we noticed that $C_{J,N}$ is
  positive in some range 
of N;we can have a stable bolt as well and therefore we can nucleate long-lived bolt solutions.\\
 
  From the expressions of action it is found that it will have three zeroes at
\begin{eqnarray} 
\nonumber{(r_+)}_0^2 = (a^2+N^2) 
\end{eqnarray} 
and also at
\begin{eqnarray}
\nonumber {(r_+)}_1^2=\left[(a^2+l^2)+\sqrt{(a^2+l^2)^2-4(N^2-a^2)(3N^2-l^2)}\right]/2 
\end{eqnarray}
\begin{eqnarray}
\nonumber {(r_+)}_2^2=\left[(a^2+l^2)-\sqrt{(a^2+l^2)^2-4(N^2-a^2)(3N^2-l^2)}\right]/2
\end{eqnarray}

The action will be positive for ${r_+}^2<{(r_+)}_1^2$ and negative for ${r_+}^2>{(r_+})_1^2$.The action I as a function of$ r_+$ is plotted
in Fig6 . At small $ r_+$ temperature is low and action is positive whereas at large $ r_+$, temperature is high and the action is negative
. Thus Kerr-Bolt-Ads state will be preferred at high temperature.\\

{\bf \large{Concluding Remarks}}:\\
Anti-de Sitter space has been regarded as of little physical significance.Extended theories of supergravity have anti-de Sitter space as

their ground or most symmetric state. Asymptotically anti-deSitter solutions are stable even though the potentials that appear in the theories
are unbounded below.So, study of quantum mechanical and thermodynamical properties of black-holes are important in anti-de Sitter spaces.\\
The purpose of our paper is to consider thermodynamical properties of black holes in Kerr-Bolt-Ads spacetime.When the Nut charge is non-
vanishing, the presence of rotation does not admit the existence of regular spacetime solutions unless a bolt is also present.
One of the unusual results is that entropy is not positive for all values of the parameters, shown in Fig 1.
Angular velocity $\Omega$ decreases with decreasing J once M becomes sufficiently large and then vanishes. In some range of physical
 parameters N, specific heat becomes positive and we can have a stable bolt solutions. It was shown in [7] that there is a critical 
temperature $T_c$ at  which tharmal radiation is unstable to the formation of a schwarzschild black hole. For$ T> {T_c}$, there are
two values of the black hole mass at which Hawking radiation can be in equilibrium with the thermal radiation of the background. In
[22] a phase transition from TN-Ads to TB-Ads was occured. In our case there is a generalisation of TB-Ads to Kerr-TB-Ads spacetime.
At high temperature, the action being negative, Kerr-Bolt-Ads state will be preferred.\\

{\bf\large{Acknowledgement}}:\\ T.G wishes to thank C.S.I.R for financial support and also Prof.Partha Mitra,Theory division
Saha Institute Of Nuclear Physics for his comments.

\begin{figure}
\begin{center}
\hspace{10cm}
\epsfxsize=5in
\epsfbox{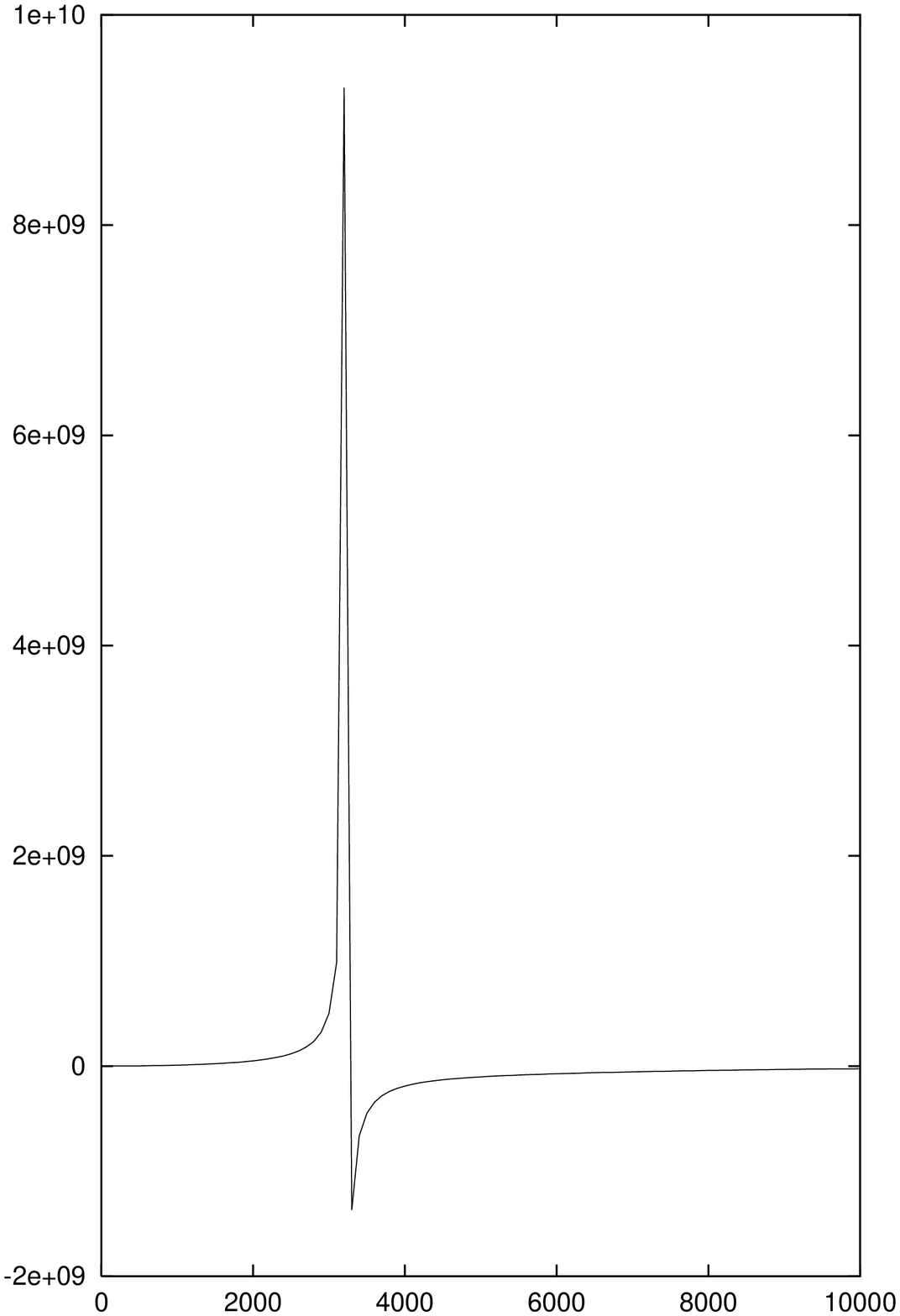}
\label{Figure:1}
\caption{ S versus N}
\end{center}
\end{figure}

\begin{figure}
\begin{center}
\hspace{10cm}
\epsfxsize=5in
\epsfbox{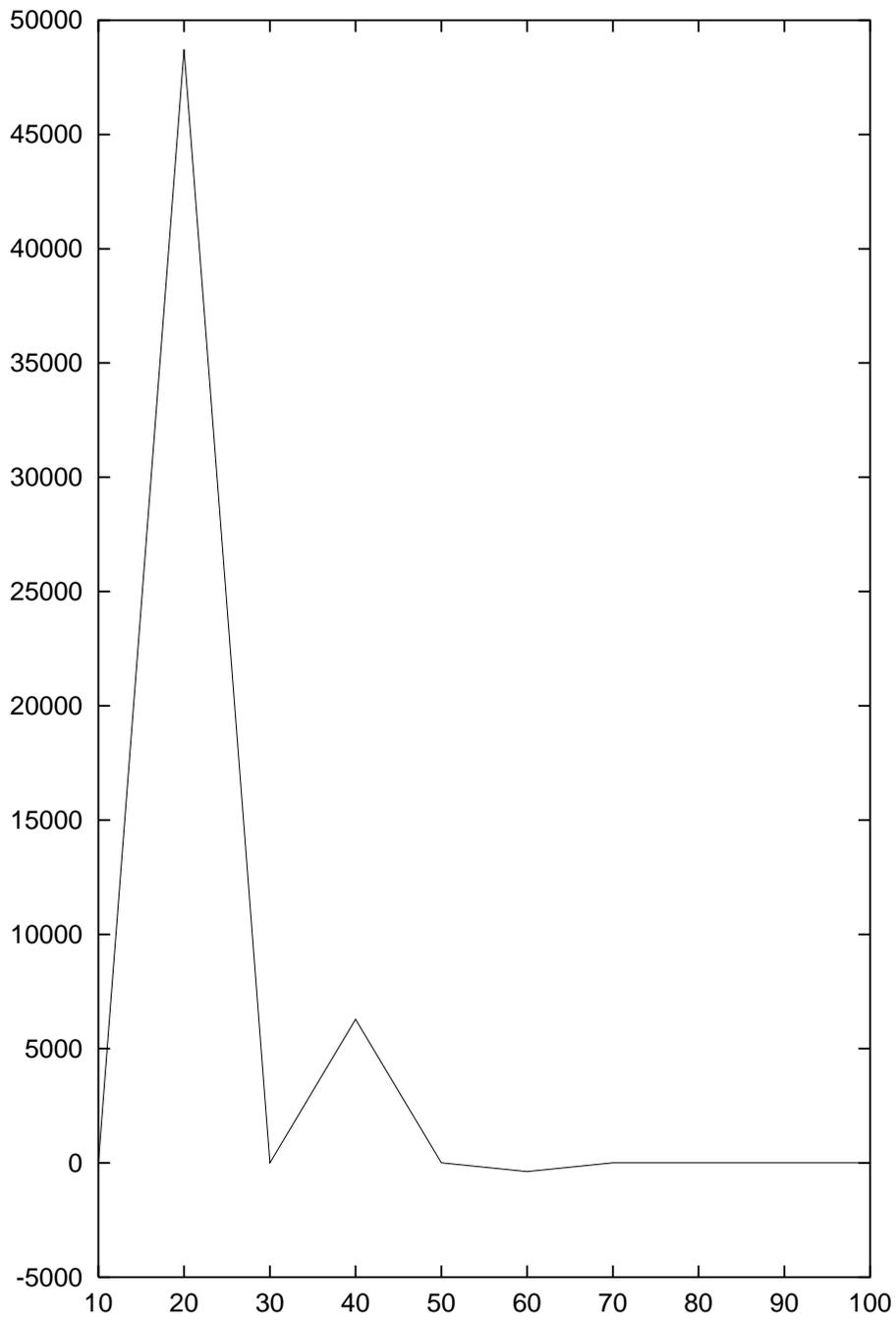}
\label{fig:2}
\caption{$\Omega$ versus J}
\end{center}
\end{figure}

\begin{figure}
\begin{center}
\hspace{10cm}
\epsfxsize=5in
\epsfbox{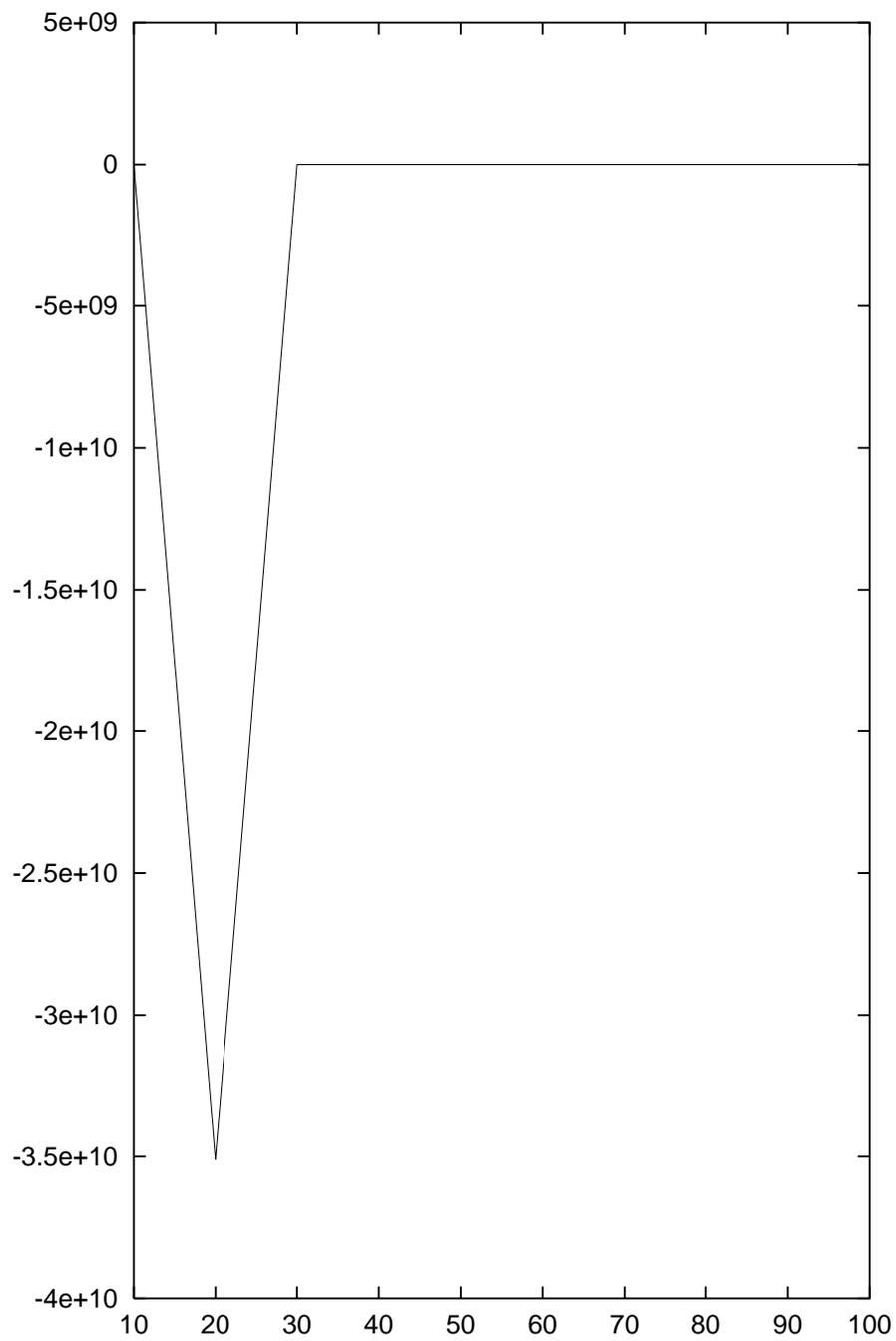}
\label{fig:3}
\caption{ $\Omega$ versus J;$\Omega$ is not positive for large M}
\end{center}
\end{figure}

\begin{figure}
\begin{center}
\hspace{10cm}
\epsfxsize=5in
\epsfbox{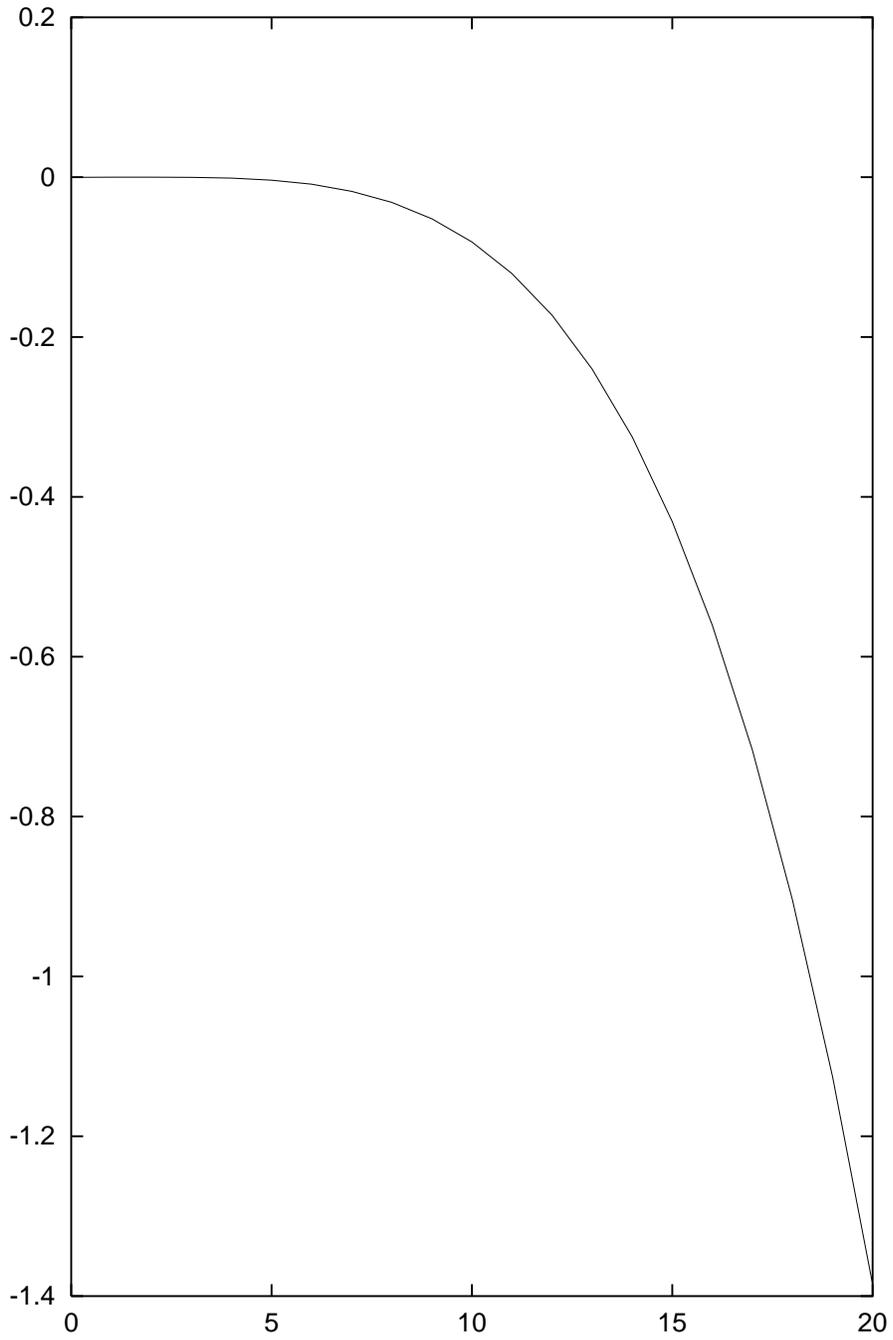}
\label{fig:4}
\caption{ $C_{J,N}$ is plotted as a function of N;$C_{J,N}$ is not positive in this range of N.}
\end{center}
\end{figure}

\begin{figure}
\begin{center}
\hspace{10cm}
\epsfxsize=5in
\epsfbox{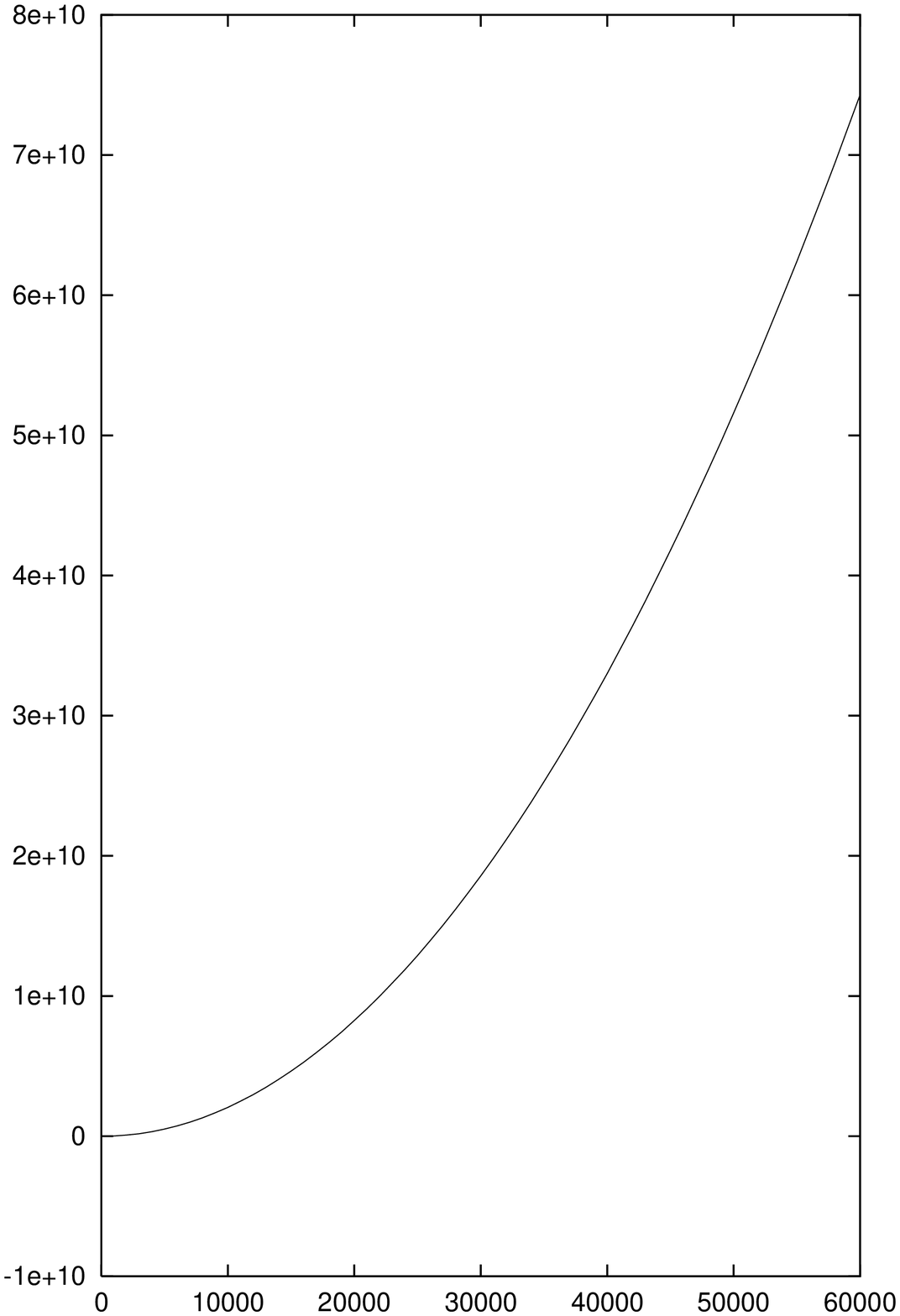}
\label{fig:5}
\caption{ $C_{J,N}$ is plotted as afunction of N.$C_{J,N}$ is positive in some range of N.}
\end{center}
\end{figure}

\begin{figure}
\begin{center}
\hspace{10cm}
\epsfxsize=5in
\epsfbox{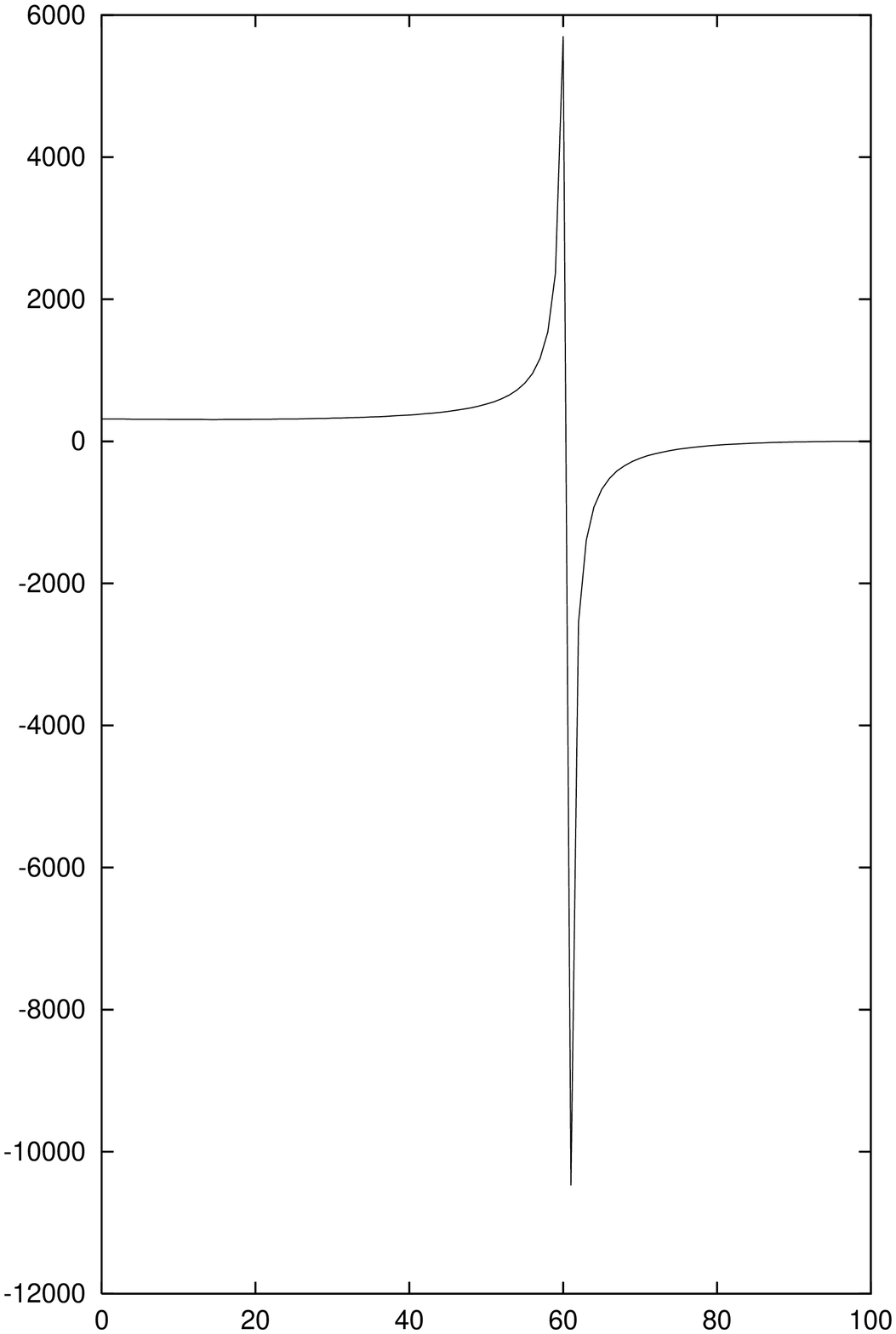}
\label{fig:6}
\caption{Action I is plotted as a function of $r_+$}
\end{center}
\end{figure}


\begin{thebibliography}{99}
\bibitem{1} J.D.Bekenstein, Phys.Rev.D {\bf 7}, 2333 (1973)\\
\bibitem{2}S.W.Hawking, Nature (London) {\bf 248}, 30 (1974); 
 Commun.Math.Phys.{\bf43},199 (1975).\\
\bibitem{3}J.Maldacena, Adv.Theor.Math.Phys. {\bf 2}, 231 (1998). \\
\bibitem{4}E.Witten, Adv.Theor.Math.Phys. {\bf 2}, 253 (1998)  \\
\bibitem{5}S.S.Gubser,I.R.Klebanov,and A.M.Polyakov, Phys.Lett.B {\bf 428}, 105 (1998).\\
\bibitem{6}O.Aharnoy, S.S.Gubser,J.Maldacena,H.Ooguri,and Y.Oz, hep-th/9905111\\
\bibitem{7}C.S.Peca and P.S.Jose Lemos, Phys.Rev.D {\bf 59}, 124007 (1999).\\
\bibitem{8}P.Mitra,Phys.Lett.B {\bf 459},119 (1999),\\
\bibitem{9} A.Chamblin,{\it et.al.}, Phys.Rev.D {\bf{60}}, 104026 (1999).\\
\bibitem{10}S.W.Hawking and D.N.Page, Commun.Math.Phys. {\bf{87}}, 577 (1983) .\\
\bibitem{11}M.Cvetic and S.S.Gubser, JHEP {\bf{ 04}}, 024, (1999).\\
\bibitem{12}S.W Hawking, {\it et.al.}, Phys.Rev.D {\bf 59}, 064005 (1999).\\
\bibitem{13}S.W Hawking, {\bf \it Talk given at Strings 99}, Posdam (Germany), July 19-24 (1999).\\
\bibitem{14}P.C.W.Davies, Class.Quant.Grav {\bf{ 6}}, 1909 (1989).\\
\bibitem{15}D.S.Berman and M.K.Parikh, Phys.Lett.B {\bf{ 463}}, 168 (1999).\\
\bibitem{16}L.Smarr, Phys.Rev.Lett {\bf{30}},71 (1973).\\
\bibitem{17}R.B.Mann, Phys.Rev.D {\bf{60}},104047 (1999).\\
\bibitem{18}S.W.Hawking, C.J.Hunter and  D.N.Page, Phys.Rev.D {\bf{59}}, 044033 (1999).\\
\bibitem{19}R.B.Mann, Phys.Rev.D {\bf{61}},084013 (2000).\\
\bibitem{20}G.W.Gibbons and M.Perry, Phys.Rev.D {\bf{22}},313 (1980).\\
\bibitem{21}M.H.Dehghani and R.B.Mann, Phys.Rev.D {\bf{64}}, 044003 (2001).\\
\bibitem{22}A.Chamblin, {\it et.al.}, Phys.Rev.D {\bf{59}}, 064010 (1999).\\
\end{thebibliography}
\end{document}